\documentstyle[12pt]{article}

\def\beq{\begin{equation}}
\def\eeq{\end{equation}}
\def\bea{\begin{eqnarray}}
\def\eea{\end{eqnarray}}
\def\sss{\scriptscriptstyle}

\def\bd{B_d^0}
\def\bdbar{{\overline{B_d^0}}}
\def\bs{B_s^0}
\def\bsbar{{\overline{B_s^0}}}
\def\barp{{\raise.35ex\hbox
{${\sss (}$}}---{\raise.35ex\hbox{${\sss )}$}}}
\def\bdbarp{\hbox{$B_d$\kern-1.4em\raise1.4ex\hbox{\barp}}}
\def\bsbarp{\hbox{$B_s$\kern-1.4em\raise1.4ex\hbox{\barp}}}
\def\ks{K_{\sss S}}
\def\kbar{{\overline{K^0}}}
\def\roughly#1{\mathrel{\raise.3ex\hbox
{$#1$\kern-.75em\lower1ex\hbox{$\sim$}}}}
\def\lsim{\roughly<}

\def\prd#1#2#3{{\it Phys.\ Rev.} {\bf D#1}, #3 (19#2)}
\def\nn{\nonumber}
\def\Ptilde{{\tilde P}}
\def \zpc#1#2#3{{\it Z.~Phys.,} {\bf C#1} (19#2) #3}
\def \plb#1#2#3{{\it Phys.~Lett.,} {\bf B#1} (19#2) #3}

\def \prl#1#2#3{{\it Phys.~Rev.~Lett.,} {\bf #1} (19#2) #3}
\def \prd#1#2#3{{\it Phys.~Rev.,} {\bf D#1} (19#2) #3} 
\def \npb#1#2#3{{\it Nucl.~Phys.}, {\bf B#1} (19#2) #3} 
\def\ijmp#1#2#3{{\it Int.\ J.\ Mod.\ Phys.} {\bf A#1} (19#2) #3}
\def \stone{{\it B Decays}, edited by S. Stone (World Scientific, Singapore,
1994)}
\begin{document}
\setlength{\baselineskip}{20pt}

\begin{flushright}
{\bf hep-ph/9904311} \\
YUMS 99-007\\
UdeM-GPP-TH-99-59 \\
KEK-TH-621 \\
\end{flushright}

\begin{center}
\bigskip

%{\Large \bf Searching for New Physics through CP-phase 
%            $\beta~(\equiv \phi_1)$}\\ 
{\Large \bf Searching for New Physics in the $b\to d$ FCNC}\\
\vskip1truemm
{\Large \bf using $\bs(t)\to\phi\ks$ and $\bd(t) \to K^0 \kbar$}\\ 
\bigskip

C. S. Kim$^{a,}$\footnote{kim@cskim.yonsei.ac.kr,~~
http://phya.yonsei.ac.kr/\~{}cskim/ },~~ 
D. London $^{b,}$\footnote{london@lps.umontreal.ca}~~ 
and~~ T. Yoshikawa$^{c,}$\footnote{JSPS Research Fellow,~~
yosikawa@acorn.kek.jp}
\end{center}

%\medskip

\begin{flushleft}
~~~~~~~~~~~$a$: {\it Department of Physics, Yonsei University, 
Seoul 120-749, Korea}\\
~~~~~~~~~~~$b$: {\it Laboratoire Ren\'e J.-A. L\'evesque, 
Universit\'e de Montr\'eal,}\\
~~~~~~~~~~~~~~~{\it C.P. 6128, succ. centre-ville, Montr\'eal, QC,
Canada H3C 3J7}\\
~~~~~~~~~~~$c$: {\it Theory Group, KEK, Tsukuba, Ibaraki 305-0801, Japan}
\end{flushleft}

\begin{center}

\bigskip
(\today)

\bigskip 

{\bf Abstract}

\end{center}

\begin{quote}
  Time-dependent measurements of the decays $\bs(t)\to\phi\ks$ and
  $\bd(t) \to K^0 \kbar$ can be used to compare the weak phases of
  $\bd$-$\bdbar$ mixing and the $t$-quark contribution to the $b\to d$
  penguin. Since these two phases are equal in the standard model, any
  discrepancy would be a sign of new physics, specifically in the
  $b\to d$ flavour-changing neutral current (FCNC). The method can be
  applied to other pairs of decays, such as $\bs(t)\to J/\Psi \ks$ and
  $\bd(t)\to J/\Psi \pi^0$.
\end{quote}
\newpage

\section{Introduction}

Over the past decade or so, many methods have been proposed for
obtaining the three interior angles of the unitarity triangle,
$\alpha$, $\beta$ and $\gamma$. In the near future these CP phases
will be measured in a variety of experiments at $B$-factories, HERA-B,
and hadron colliders. As always, the hope is that these measurements
will reveal the presence of physics beyond the standard model (SM).

The CP angles are typically extracted from CP-violating rate
asymmetries in $B$ decays \cite{BCPasym}. New physics, if present,
will affect these asymmetries principally through new contributions to
loop-level processes. Most asymmetries involve the tree-level decays
of neutral mesons ($\bd$ or $\bs$), and the new physics can enter into
$B^0$-${\overline{B^0}}$ mixing \cite{NPBmixing}. However, it is
occasionally the case that penguin contributions are involved, and the
new physics can enter here as well \cite{GroWorah}.

The canonical decay modes for measuring $\alpha$ and $\beta$ are
$\bd(t) \to \pi^+\pi^-$ and $\bd(t) \to J/\Psi \ks$, respectively.
There have been many proposals for measuring the angle $\gamma$. One
of these, accessible at asymmetric $e^+e^-$ $B$-factories, involves
the CP asymmetry in $B^\pm \to D K^\pm$ \cite{BDK}. Another method,
which is more appropriate to hadron colliders, uses $\bs(t) \to
D_s^\pm K^\mp$ \cite{BsDsK}.

How will we know whether or not new physics is present? One obvious
way is if the three angles do not add up to $180^\circ$.
Unfortunately, if the CP-angles are obtained in the conventional ways
described above, $B$-factories can never find $\alpha+\beta+\gamma \ne
\pi$. The reason is as follows: if there is new physics in
$\bd$-$\bdbar$ mixing, the CP asymmetries in $\bd(t) \to \pi^+\pi^-$
and $\bd(t) \to J/\Psi \ks$ will both be affected, but in opposite
ways: instead of measuring $\alpha$ and $\beta$, the true (SM)
CP-angles, one will extract $\tilde{\alpha } = \alpha - \theta_{\sss
  NP}$ and $\tilde{\beta }= \beta + \theta_{\sss NP}$, where
$\theta_{\sss NP}$ is the new-physics phase \cite{NirSilv}. On the
other hand, since the measurement of $\gamma$ does not involve $\bd$
decays, it will be unaffected by new physics. Thus, even in the
presence of new physics, one will still find $\tilde{\alpha} +
\tilde{\beta} + \gamma = \pi$.

$B$-factories must therefore find other ways of testing for the
presence of new physics \cite{newphysics}. The most common method is
simply to compare the unitarity triangle constructed from measurements
of the angles with that constructed from independent measurements of
the sides. If there is a discrepancy, one can then deduce that new
physics is present, probably in $\bd$-$\bdbar$ mixing. The problem
here is that there are large theoretical errors in obtaining the
length of the sides of the unitarity triangle from the experimental
data. Because of this, the presently-allowed region for the unitarity
triangle is still rather large \cite{AliLondon}. Thus, if
$\theta_{\sss NP}$ is relatively small, one may still find agreement
in the unitarity triangles constructed from the angles and sides.
Furthermore, even if there is a discrepancy, one isn't sure whether
new physics really is present --- it may simply be that the errors on
the theoretical input quantities have been underestimated.  The point
here is that we would really like to find a method of directly probing
new physics in $\bd$-$\bdbar$ mixing.

Note that there are a variety of ways of testing for new physics in
$\bs$-$\bsbar$ mixing, or, more precisely, in the $b\to s$
flavour-changing neutral current (FCNC). For example, above we
mentioned two methods for obtaining $\gamma$: via $B^\pm \to D K^\pm$
and $\bs(t) \to D_s^\pm K^\mp$. If there is a discrepancy in the value
of $\gamma$ obtained from these two decays, this would be a direct
indication of new physics in $\bs$-$\bsbar$ mixing. A second example,
somewhat different, is to measure $\beta$ using the decay $\bd(t) \to
\phi \ks$ \cite{LonSoni}. This is a pure $b\to s$ penguin decay. Thus,
if the values of $\beta$ extracted via $\bd(t) \to J/\Psi \ks$ and
$\bd(t) \to \phi \ks$ were to disagree, this would indicate the
presence of new physics in the $b\to s$ penguin, i.e.\ in the $b\to s$
FCNC. (There might also be new physics in $\bd$-$\bdbar$ mixing, but
the effect would be the same for the two decays.) Since new physics
which affects $\bs$-$\bsbar$ mixing is also likely to affect the $b\to
s$ penguin, in some sense this is a way of probing new physics in
$\bs$-$\bsbar$ mixing without actually using $\bs$ mesons. Finally, a
third example involves the CP asymmetry in $\bs(t) \to J/\Psi \phi$.
To a good approximation, this asymmetry vanishes in the SM, so that a
nonzero value would be clear evidence of new physics, specifically in
$\bs$-$\bsbar$ mixing.

Although there are many ways of getting at new physics in the $b\to s$
FCNC, to date no method has been suggested which directly tests for
new physics in the $b\to d$ FCNC\footnote{In Ref.~\cite{Dalitz} it was
  claimed that the study of the Dalitz plot of $\bd(t)\to
  \pi^+\pi^-\pi^0$ decays allows one to cleanly perform such tests.
  However, it has since been shown that this particular point is in
  error, see Ref.~\cite{LSS}.}. In this Letter, we propose a method
for doing just this. Essentially, the technique compares the weak
phase of $\bd$-$\bdbar$ mixing with that of the $t$-quark contribution
to the $b\to d$ penguin. In the SM, these phases are the same, since
they involve the same Cabibbo-Kobayashi-Maskawa (CKM) \cite{CKM}
matrix elements $V_{tb}^* V_{td}$.  However, if there is new physics
in the $b\to d$ FCNC, there may be a discrepancy. The method involves
the decay $\bs \to \phi\ks$, along with its quark-level
flavour-$SU(3)$ counterpart, $\bd \to K^0 \kbar$. Our test of new
physics in the $b\to d$ FCNC is not entirely clean -- it involves some
theoretical input. However, the assumption we make is reasonably
well-motivated, and so this may provide a first direct probe for new
physics in the $b\to d$ FCNC.  We also discuss a variation of this
method involving the decays $\bs \to J/\Psi \ks$ and $\bd \to J/\Psi
\pi^0$.

\section{$\bs(t)\to\phi\ks$ and $\bd(t) \to K^0 \kbar$}

When discussing the weak phases probed in various CP asymmetries, it
is convenient to use approximate Wolfenstein parametrization of the
CKM matrix \cite{Wolfenstein}, in which only $V_{td}$ and $V_{ub}$
have significant non-zero phases. The CKM phases in the unitarity
triangle are then $\beta = {\rm Arg}(V_{td}^*)$ and $\gamma = {\rm
  Arg}(V_{ub}^*)$, with $\alpha$ defined to be $\pi - \beta - \gamma$.
We will use this parametrization throughout the paper.

We begin by considering the decay $\bs\to\phi\ks$. The CP asymmetry
in $\bs(t)\to\phi\ks$ measures the relative phases of the two
amplitudes ($\bs\to\phi\ks$) and
($\bs\to\bsbar$)($\bsbar\to\phi\ks$). To begin with, let us assume
that there is no new physics. $\bs\to\phi\ks$ is a pure penguin
decay, which at the quark level takes the form ${\bar b} \to {\bar d}
s {\bar s}$.  Suppose first that this decay is dominated by an
internal $t$-quark, in which case the CKM matrix-element combination
involved is $V_{tb}^* V_{td}$. Since $\bs$-$\bsbar$ mixing involves
$(V_{tb}^* V_{ts})^2$, the CP asymmetry probes
\beq
{\rm Arg} \left[ { V_{tb}^* V_{td} \over (V_{tb}^* V_{ts})^2 V_{tb}
    V_{td}^* } \right] = -2 \beta ~.
\eeq
Thus, within the SM, if the $t$-quark dominates the $b\to d$ penguin,
one expects the CP asymmetry in $\bd(t) \to J/\Psi \ks$ to be equal to
that in $\bs(t) \to \phi\ks$ \cite{LonPeccei}.

If there is new physics, the CP asymmetry in $\bs(t)\to\phi\ks$ can
be affected in two ways: there may be new contributions to
$\bs$-$\bsbar$ mixing (the $b\to s$ FCNC) and/or to the penguin decay
$\bs\to\phi\ks$ (the $b\to d$ FCNC). As discussed previously, new
physics in $\bs$-$\bsbar$ mixing can be discovered independently, for
example by comparing the CP asymmetries in $B^\pm \to D K^\pm$ and
$\bs(t) \to D_s^\pm K^\mp$. If, after taking this into account, there
is still a discrepancy in the value of $\beta$ as extracted from the
CP asymmetries in $\bd(t) \to J/\Psi \ks$ and $\bs(t) \to\phi\ks$,
this will indicate the presence of new physics in the $b\to d$ FCNC.

Note also that one can perform a similar analysis with the related
decay $\bd \to K^0 \kbar$ \cite{LonPeccei}. (At the quark level, the
decays are identical, save for the flavour of the spectator quark; at
the meson level there is a difference since here there are two
pseudoscalars in the final state, while the $\bs$ decay has a vector
and a pseudoscalar.) Under the same assumptions as above, the CP
asymmetry in $\bd(t) \to K^0 \kbar$ measures
\beq
{\rm Arg} \left[ { V_{tb}^* V_{td} \over (V_{tb}^* V_{td})^2 V_{tb}
    V_{td}^* } \right] = 0 ~.
\eeq
Thus, a non-zero CP asymmetry in this mode would indicate the presence
of new physics in the $b\to d$ FCNC.

However, there is a problem with the above analysis. Theoretical
estimates suggest that the $b\to d$ penguin is {\it not} dominated by
an internal $t$-quark. On the contrary, the $u$- and $c$-quark
contributions can be substantial, perhaps even as large as 20\%--50\%
of the $t$-quark contribution \cite{ucquark}. If this is the case,
then the CP asymmetry in $\bs(t)\to\phi\ks$ no longer cleanly probes
the angle $\beta$.  Instead, there is now ``penguin pollution'' from
the $u$- and $c$-quark contributions to the $b\to d$ penguin, and the
result depends on (unknown) hadronic quantities such as the strong
phases and the relative sizes of the various penguin contributions.
Thus, if one wants to detect new physics in the $b\to d$ FCNC, it is
necessary to deal with this penguin pollution. As we show below, this
can be done by combining information from both $\bs(t)\to\phi\ks$ and
$\bd \to K^0 \kbar$.

The $B_s$ and $B_d$ systems differ in that the width difference
between the light and heavy $B_s$ eigenstates may be measurable, which
is not the case for the $B_d$ system. In the presence of a
non-negligible width difference, the expressions for the
time-dependent decays of $B_s$ mesons are \cite{Dunietz}
\bea
\label{Bsrates}
\Gamma(\bs(t) \to f) & = & |A_f|^2 |g_+(t)|^2 + |{\bar A}_f|^2 |g_-(t)|^2 
\nn \\
& ~ & \quad
+ 2  \left[ {\rm Re} (A_f^* {\bar A}_f) {\rm Re} (g_-(t) g_+^*(t))
- {\rm Im} (A_f^* {\bar A}_f) {\rm Im} (g_-(t) g_+^*(t)) \right] ~, \nn \\
\Gamma(\bsbar(t) \to f) & = & |{\bar A}_f|^2 |g_+(t)|^2 + |A_f|^2 |g_-(t)|^2 
\nn \\
& ~ & \quad
+ 2  \left[ {\rm Re} (A_f^* {\bar A}_f) {\rm Re} (g_-(t) g_+^*(t))
+ {\rm Im} (A_f^* {\bar A}_f) {\rm Im} (g_-(t) g_+^*(t)) \right] ~,
\eea
with 
\bea
|g_+(t)|^2 + |g_-(t)|^2 & = & {1\over 2} \left( e^{-\Gamma_{\sss L} t} +
                                 e^{-\Gamma_{\sss H} t} \right)~, \nn \\
|g_+(t)|^2 - |g_-(t)|^2 & = & e^{-\Gamma t} \cos\Delta m t  ~, \nn \\
{\rm Re} [g_-(t) g_+^*(t)] & = & {1\over 4} \left( e^{-\Gamma_{\sss L} t} -
e^{-\Gamma_{\sss H} t} \right) ~, \nn \\
{\rm Im} [g_-(t) g_+^*(t)] & = & {1\over 2} e^{-\Gamma t} \sin\Delta m t ~.
\eea
In the above, $\Gamma_{\sss L}$ and $\Gamma_{\sss H}$ are the widths
of the light and heavy $B$-states, respectively, and $ \Gamma \equiv
(\Gamma_{\sss L} + \Gamma_{\sss H}) / 2$. (Note also that we have
assumed that the weak phase in $\bs$-$\bsbar$ mixing is zero, which
holds within the SM.)

Thus, the time-dependent measurements of $B_s$ decay rates allow one
to obtain the following four functions of the decay amplitudes:
\beq
|A_f|^2 + |\bar{A}_f|^2 ~~,~~ 
|A_f|^2 - |\bar{A}_f|^2 ~~,~~ 
{\rm Re}\left( A_f^* \bar{A}_f \right) ~~,~~
{\rm Im}\left( A_f^* \bar{A}_f \right).
\label{Aquantities}
\eeq
If the width difference in the $B_s$ system turns out to be small,
then ${\rm Re} [g_-(t) g_+^*(t)] \simeq 0$, which appears to imply
that one cannot obtain the quantity ${\rm Re}\left( A_f^* \bar{A}_f
\right)$. However, this is not true. In fact, the above four functions
are not independent, due to the equality
\beq
|A_f|^2 |{\bar A}_f|^2 = [{\rm Re} (A_f^* {\bar A}_f)]^2 + [{\rm Im} (A_f^*
{\bar A}_f)]^2 ~.
\eeq
Thus, even if the width difference in the $B_s$ system is not
measurable, we can still obtain each of the four quantities in
Eq.~(\ref{Aquantities}), except that the {\it sign} of ${\rm Re}\left(
A_f^* \bar{A}_f \right)$ is undetermined. The lack of knowledge of
this sign simply leads to additional possible solutions (discrete
ambiguities), one for each sign of ${\rm Re}\left( A_f^* \bar{A}_f
\right)$. A measurable width difference allows the determination of
this sign, thereby reducing discrete ambiguities. In what follows, we
assume that the width difference is measurable, i.e.\ that the sign of
${\rm Re}\left( A_f^* \bar{A}_f \right)$ is known. If this sign cannot
be determined, the method is still valid, but extra solutions are
possible.

Now, consider again the process $\bs\to\phi\ks$. At the level of
quark diagrams, there are several contributions to this decay: (i) the
ordinary gluonic penguin, $\Ptilde$, (ii) the Zweig-suppressed gluonic
decay $\Ptilde_1$, in which the gluon essentially hadronizes into the
$\phi$, (iii) the electroweak penguin $\Ptilde_{\sss EW}$, and (iv)
the colour-suppressed electroweak penguin $\Ptilde_{\sss EW}^{\sss
  C}$.  (The tildes on the amplitudes indicate a $\bs$ decay.) We
therefore write the amplitude schematically as
\bea
A_s^{\phi} &\equiv & A\left(B_s\rightarrow \phi\ks\right) = 
{1\over \sqrt{2}} \left( \Ptilde + \Ptilde_1 + \Ptilde_{\sss EW} 
+ \Ptilde_{\sss EW}^{\sss C} \right) ~.
\label{phiksamp}
\eea
(The factor $1/\sqrt{2}$ is included due to the presence of the
$\ks$.)  Of these, the gluonic penguins receive contributions from
internal $u$, $c$ and $t$-quarks, while the electroweak penguins are
$t$-quark dominated.

Any contribution to the $b\to d$ penguin can be written generically as 
\bea
P = \sum_{q=u,c,t} V_{qb}^* V_{qd} P_q & = & V_{cb}^* V_{cd} (P_c - P_u) + 
V_{tb}^* V_{td} (P_t - P_u) \nn \\
& \equiv & {\cal P}_{cu} e^{i\delta_c} + {\cal P}_{tu} e^{i\delta_t}e^{-i\beta} ~.
\eea
In the first line we have used the unitarity of the CKM matrix to
eliminate the $u$-quark contribution, and in the second we have
explicitly separated out the weak and strong phases and absorbed the
magnitudes $|V_{cb}^* V_{cd}|$ and $|V_{tb}^* V_{td}|$ into the
definitions of ${\cal P}_{cu}$ and ${\cal P}_{tu}$, respectively.

Applying this to $\bs\to\phi\ks$, Eq.~(\ref{phiksamp}) becomes
\beq
A_s^\phi = {1\over\sqrt{2}} \left( \tilde{\cal P}_{cu} e^{i{\tilde\delta}_c}
          + \tilde{\cal P}_{tu} e^{i{\tilde\delta}_t}e^{-i\beta } \right),
\eeq
where $\tilde{\cal P}_{cu}$ and $\tilde{\cal P}_{tu}$ are real and
taken to be positive, and
\bea 
\tilde{\cal P}_{cu} e^{i{\tilde\delta}_c} & = & \Ptilde_c -
\Ptilde_u + \Ptilde_{1,c} - \Ptilde_{1,u} ~, \nn \\
\tilde{\cal P}_{tu} e^{i{\tilde\delta}_t} e^{-i\beta } & = & 
\Ptilde_t - \Ptilde_u + \Ptilde_{1,t} - \Ptilde_{1,u} +
\Ptilde_{\sss EW} + \Ptilde_{\sss EW}^{\sss C} ~.
\label{Ptildedefs}
\eea

With this expression for the $\bs\to\phi\ks$ amplitude, the
measurements of the quantities in Eq.~(\ref{Aquantities}) give
\bea
\tilde{X} &\equiv & \frac{1}{2}
            \left( 
                \left| A_s^{\phi}\right|^2 
              + \left| \bar{A}^{\phi}_s \right|^2 
            \right) = 
{1\over2} \left(
            \tilde{\cal P}_{cu}^2 +\tilde{\cal P}_{tu}^2 
      + 2 \tilde{\cal P}_{cu} \tilde{\cal P}_{tu} \cos{\tilde\Delta} \cos\beta \right) ~,
\label{XP} \\
\tilde{Y} &\equiv & \frac{1}{2}
            \left( 
                \left| A_s^{\phi}\right|^2 
              - \left| \bar{A}^{\phi}_s \right|^2 
            \right) = {1\over2} \left( - 2 \tilde{\cal P}_{cu}
                      \tilde{\cal P}_{tu} \sin{\tilde\Delta} \sin\beta \right) ~,
\label{YP} \\
\tilde{Z}_R &\equiv & Re\left( A_s^{\phi*}
                                  \bar{A}^{\phi}_{s } \right) 
            =  {1\over2} \left( 
- \tilde{\cal P}_{cu}^2 - \tilde{\cal P}_{tu}^2 \cos2\beta
      - 2 \tilde{\cal P}_{cu} \tilde{\cal P}_{tu} \cos{\tilde\Delta}\cos\beta \right) ~,
\label{ZPR} \\
\tilde{Z}_I &\equiv & Im\left( A_s^{\phi*}
                                  \bar{A}^{\phi}_{s } \right) 
            = {1\over2} \left( - \tilde{\cal P}_{tu}^2 \sin2\beta
      - 2 \tilde{\cal P}_{cu} \tilde{\cal P}_{tu} \cos{\tilde\Delta}\sin\beta \right) ~, 
\label{ZPI} 
\eea
where ${\tilde\Delta}\equiv {\tilde\delta}_c - {\tilde\delta}_t$.
(Note: in the above we have assumed that there is no new physics in
$\bs$-$\bsbar$ mixing. As mentioned previously, the presence of such
new physics can be independently determined. We will discuss the case
of new physics in $\bs$-$\bsbar$ mixing further on.)

Examining the above equations, we note that there are four unknown
parameters ($\tilde{\cal P}_{cu}$, $\tilde{\cal P}_{tu}$,
${\tilde\Delta}$ and $\beta$), but only three independent
measurements. Thus, we cannot solve for the unknowns. In particular,
we see that we cannot obtain $\beta$ in the presence of penguin
pollution.

However, progress can be made if we also consider the decay $\bd \to
K^0 \kbar$, which is similar to $\bs\to\phi\ks$. (At the quark
level, they differ only in the flavour of the spectator quark.) For
this decay, contributions come only from the ordinary gluonic penguin
$P$ and the colour-suppressed electroweak penguin $P_{\sss EW}^{\sss
C}$ (amplitudes without tildes indicate $\bd$ decays):
\beq
A_d^{\sss KK} = A\left( \bd \to K^0 \kbar\right) = P + P_{\sss EW}^{\sss C} ~.
\eeq
Analogous to the $\bs\to\phi\ks$ amplitude, we can write
\beq
A_d^{\sss KK} = {\cal P}_{cu} e^{i\delta_c}
          + {\cal P}_{tu} e^{i\delta_t}e^{-i\beta }, 
\eeq
where
\bea 
{\cal P}_{cu} e^{i\delta_c} & = & P_c - P_u ~, \nn \\
{\cal P}_{tu} e^{i\delta_t} e^{-i\beta} & = & 
P_t - P_u + P_{\sss EW}^{\sss C} ~.
\label{Pdefs}
\eea

In the $\bd$ system, the width difference is negligible, so that only
three quantities can be obtained from time-dependent measurements:
\bea
{X} &\equiv & \frac{1}{2}
            \left( 
                \left| A_d^{\sss KK}\right|^2 
              + \left| \bar{A}^{\sss KK}_d \right|^2 
            \right) = 
            {\cal P}_{cu}^{2} +{\cal P}_{tu}^{2} 
      + 2 {\cal P}_{cu} {\cal P}_{tu} \cos\Delta \cos\beta ~,
\label{XXP} \\
{Y} &\equiv & \frac{1}{2}
            \left( 
                \left| A_d^{\sss KK}\right|^2 
              - \left| \bar{A}^{\sss KK}_d \right|^2 
            \right) =  - 2 {\cal P}_{cu}
                     {\cal P}_{tu} \sin \Delta \sin\beta ~,
\label{YYP} \\
{Z} &\equiv & Im\left( e^{-2i \tilde{\beta }} A_d^{\sss KK*}
                                  \bar{A}^{\sss KK}_{d } \right) \nn \\
&~& \qquad\qquad    =  - {\cal P}_{cu}^{2} \sin 2\tilde{\beta } 
            -  {\cal P}_{tu}^{2} \sin(2 \tilde{\beta } - 2 \beta ) 
      - 2 {\cal P}_{cu} {\cal P}_{tu} \cos\Delta \sin(2 \tilde{ \beta } - \beta ) ~,
\label{ZZZP} 
\eea
where $\Delta \equiv \delta_c - \delta_t$. Since we are allowing for
the possibility of new physics in the $b\to d$ FCNC, we have
explicitly denoted the weak phase of $\bd$-$\bdbar$ mixing as
$\tilde\beta$, which may be different from the weak phase of the $b\to
d$ penguin $\beta$.

It is reasonable to assume that the mixing phase $\tilde\beta$ will be
measured independently via $\bd(t) \to J/\Psi \ks$. Even so, we are
still left with three equations in four unknowns (${\cal P}_{cu}$,
${\cal P}_{tu}$, $\Delta$ and $\beta$), so once again we cannot solve
for $\beta$. 

However, we can reduce the number of independent parameters in the
$\bs(t)\to\phi\ks$ and $\bd(t) \to K^0 \kbar$ measurements by
making an assumption. Specifically, we assume that $r = \tilde{r}$,
where $r \equiv {\cal P}_{cu}/{\cal P}_{tu}$ and $\tilde{r} \equiv
\tilde{\cal P}_{cu}/\tilde{\cal P}_{tu}$. How good is this assumption?
{}From Eqs.~(\ref{Ptildedefs}) and (\ref{Pdefs}) we have
\bea
r & = & \left\vert { P_c - P_u \over P_t - P_u + P_{\sss EW}^{\sss C}}
\right\vert \nn\\
{\tilde r} & = & \left\vert 
{ \Ptilde_c - \Ptilde_u + \Ptilde_{1,c} - \Ptilde_{1,u} \over
    \Ptilde_t - \Ptilde_u + \Ptilde_{1,t} - \Ptilde_{1,u} 
+ \Ptilde_{\sss EW} + \Ptilde_{\sss EW}^{\sss C} } \right\vert ~.
\eea
Now, at the quark level the only difference between the $P$ and the
$\Ptilde$ amplitudes is the flavour of the spectator quark. Since this
flavour should not have a significant effect on the size of the
amplitude, for a given type of penguin contribution we can take $|P_i|
\simeq |\Ptilde_i|$.  Furthermore, we can estimate the relative sizes of
the various types of penguin contribution: $|\Ptilde_{\sss EW} /
\Ptilde_t| \simeq |\Ptilde_{1,q} / \Ptilde_q| \simeq |\Ptilde_{\sss
  EW}^{\sss C} / \Ptilde_{\sss EW}| \simeq {\bar\lambda}$, where
${\bar\lambda} \sim 20\%$. Thus, we find that
\beq
r \simeq {\tilde r} = \left\vert { P_c - P_u \over P_t - P_u} \right\vert 
\eeq
and
\beq
{r - {\tilde r} \over r} = O({\bar\lambda}) ~.
\eeq
Taking $r = \tilde{r}$ is therefore a reasonable assumption.

With the assumption that $ r = \tilde{r}$, the measurements take the form
\bea
\tilde{X} &=& {1\over 2} \, \tilde{\cal P}_{tu}^2 
                [ 1 + 2 r \cos\tilde{\Delta} \cos\beta + r^2 ] ~,
\label{XBSPK} \\
\tilde{Y} &=& {1\over 2} \, \tilde{\cal P}_{tu}^2 
                [ - 2 r \sin\tilde{\Delta} \sin\beta ] ~,
\label{YBSPK} \\
\tilde{Z}_R &=& {1\over 2} \, \tilde{\cal P}_{tu}^2 
                [ - \cos2\beta - 2 r \cos\tilde{\Delta}\cos\beta - r^2 ] ~,
\label{ZRBSPK} \\
\tilde{Z}_I &=& {1\over 2} \, \tilde{\cal P}_{tu}^2 
                [ - \sin2\beta - 2r \cos\tilde{\Delta}\sin\beta ] ~,
\label{ZIBSPK} \\
X &=& {\cal P}_{tu}^2 [ 1 + 2 r \cos{\Delta} \cos\beta + r^2 ] ~,
\label{XBDKK} \\
Y &=& {\cal P}_{tu}^2 [ - 2 r \sin{\Delta} \sin\beta ] ~,
\label{YBDKK} \\
Z &=& {\cal P}_{tu}^2 [ - \sin( 2\tilde{\beta} - 2\beta ) - 
                    2 r \cos\Delta \sin( 2\tilde{\beta} - \beta )
                    - r^2 \sin2\tilde{\beta } ] ~.
\label{ZBDKK}
\eea
Assuming that ${\tilde\beta}$ is measured in $\bd(t) \to J/\psi \ks$,
we now have six independent equations in six unknowns.

We can solve for $\beta$ as follows. First, ${\cal P}_{tu} $ and
$\tilde{\cal P}_{tu}$ are eliminated by dividing the equations as
follows:
\bea
\tilde{M} &\equiv & - \frac{\tilde{Z}_R}{\tilde{X}} =
            \frac{ \cos2\beta + 2r \cos{\tilde\Delta}\cos\beta + r^2 }
                 { 1 + 2 r \cos{\tilde\Delta}\cos\beta + r^2 }
       = - 1 + \frac{2 \sin^2\beta }{ 1 + 2 r \cos{\tilde\Delta}
                             \cos\beta + r^2 } 
\label{TMSM} \\
\tilde{N} &\equiv &  \frac{\tilde{Z}_I}{\tilde{X}} =
             \frac{ \sin2\beta + 2r \cos{\tilde\Delta}\sin\beta  }
                     { 1 + 2 r \cos{\tilde\Delta}\cos\beta + r^2 }
       = - 2 \sin\beta \frac{\cos\beta + r \cos\tilde{\delta } }
                      { 1 + 2 r \cos{\tilde\Delta}\cos\beta + r^2 } 
\label{TNSM} \\
\tilde{O} &\equiv & \frac{\tilde{Y}}{\tilde{X}} =
             \frac{ - 2 r \sin{\tilde\Delta} \sin\beta }
                      { 1 + 2 r \cos{\tilde\Delta}\cos\beta + r^2 } \\
M &\equiv & \frac{{Z}}{{X}} + \sin2\tilde{\beta}
           =\sin2\tilde{\beta} ( \frac{  2 \sin^2\beta }
                  {  1 + 2 r \cos{\Delta}\cos\beta + r^2 } ) 
           + \cos2\tilde{\beta}
                \frac{\sin2\beta + 2 r \cos\Delta \sin\beta }
                {  1 + 2 r \cos{\Delta}\cos\beta + r^2 } 
\label{M}\\
O &\equiv & \frac{{Y}}{{X}}
     =   \frac{ - 2 r \sin{\Delta} \sin\beta }
                      { 1 + 2 r \cos{\Delta}\cos\beta + r^2 }
\label{O} 
\eea
We then define
\bea
R &\equiv & - \frac{\tilde{M}+1}{\tilde{N}}
       = \frac{\sin\beta}{\cos\beta + r \cos{\tilde\Delta}} 
\label{R}
\eea 
Eliminating $\tilde{\delta}$ from Eqs.~(\ref{R}) and (\ref{TMSM}), we
then find $r^2 $ as a function of $\beta $:
\bea
r^2 = -1 + 2 \cos^2\beta + \frac{2}{\tilde{M}+1}\sin^2\beta 
         - \frac{2}{R}\sin\beta \cos\beta 
\label{r}
\eea 
And from Eq.~(\ref{M}) we have
\bea
\cos\Delta = \frac{ -2 \sin2\tilde{\beta} \sin^2\beta 
                    -2 \cos2\tilde{\beta} \sin\beta\cos\beta + M
                    + r^2 M }
                   { 2r (- M \cos\beta + \cos2\tilde{\beta}\sin\beta )}
\label{Cdel}
\eea
Finally, by inserting the expressions for $\Delta$ and $r^2 $ into
Eq.~(\ref{O}) we obtain an equation for $\beta$ in terms of
observables alone. Note that we have not used the expression for
$\tilde{Y}$ [Eq.~(\ref{YBSPK})] in the above derivation. As explained
earlier, $\tilde{Y}$ is not independent of $\tilde{X}$, $\tilde{Z}_R$
and $\tilde{Z}_I$. However, it can be used as a check to eliminate
some of the solutions, thereby reducing the discrete ambiguity.

We illustrate this solution numerically in Table~\ref{betatable}. By
choosing input values for the theoretical parameters, we can generate
the ``experimental data'' of Eqs.~(\ref{XBSPK})-(\ref{ZBDKK}). The
above method can then be used to solve for the theoretical unknowns,
and we can check that, despite the presence of multiple solutions, we
can still find that $\beta \ne {\tilde\beta}$. For the amplitudes, we
take $(\tilde{\cal P}_{tu})_{in}=1.0$ and $({\cal P}_{tu})_{in}=1.2$.
(We take these quantities to be unequal in order to account for two
things: (i) the different spectator quarks and (ii) the different
final state -- two pseudoscalars in one case, and one vector and one
pseudoscalar in the other case.) The assumed input values of $r$ and
the weak and strong phases are shown in the Table. The angles are
taken to lie in the region $0 < \beta, \Delta, {\tilde\Delta} < \pi $.

\begin{table}
\begin{center}
\begin{tabular}{|r|r r r r|r|r|r|r|r|}
\hline 
$\tilde{\beta}_{in}$ & $\beta_{in}$ & $r_{in}$ & 
$\tilde{\Delta }_{in} $ & $\Delta_{in} $ & 
$\beta $ & $ r $ & ${\tilde\Delta} $ & $\Delta $ & $ \tilde{P}_{tu}$ 
                                   \\ \hline
25 & 10 & 0.3 & 80 & 120 & 19.7 & 1.22 & 28.1 & 180.0 & 0.26 \\ 
   &    &     &    &     & 10.0 & 0.30 & 80.0 & 120.0 & 1.00 \\
   &    &     &    &     & 160.3& 1.22 & 151.9 & 0.0 & 0.26 \\
   &    &     &    &     & 170.0 & 0.30 & 100.0 & 60.0 & 1.00 \\
\hline 
25 & 40 & 0.3 & 80 & 120 & 40.0 & 0.30 & 80.0 & 120.0 & 1.00 \\
   &    &     &    &     & 57.2 & 0.65 & 35.8 & 165.9 & 0.59 \\
   &    &     &    &     & 140.0 & 0.30 & 100.0 & 60.0 & 1.00 \\
   &    &     &    &     & 122.8 & 0.65 & 144.2 & 14.1 & 0.59 \\
\hline
40 & 25 & 0.3 & 40 & 10  & 64.4  & 2.04 & 11.6 & 178.5 & 0.47 \\
  &    &      &    &     & 25.0 & 0.30 & 40.0 & 10.0 & 1.00 \\
   &   &      &    &     & 115.6 & 2,04 & 168.4 & 1.53 & 0.47 \\
    &   &     &    &     & 155.0 & 0.30 & 168.3 & 170.0 & 1.0 \\
\hline 
\end{tabular}
\end{center}
\caption{Output values of $\beta$, $r$, ${\tilde\Delta}$,  $\Delta$ and 
  $\tilde{P}_{tu}$ for given input values of $r$ and the weak and
  strong phases. We take $(\tilde{\cal P}_{tu})_{in}=1.0$ and $({\cal
    P}_{tu})_{in}=1.2$. All phase angles are given in degrees.}
\label{betatable}
\end{table}

{}From the Table, we see that $\beta$ can be extracted with a fourfold
ambiguity. However, none of these values is equal to the value of
${\tilde\beta}$. Thus, if these measurements were carried out, and
these results found, we would have unequivocal evidence of new physics
in the $b\to d$ FCNC. We would not know if it affected $\bd$-$\bdbar$
mixing, the $b\to d$ penguin, or both, but we would know with
certainty that new physics was present.

In describing this method, we have assumed that there is no new
physics in $\bs$-$\bsbar$ mixing. However, even if we include this,
the above method does not change significantly. If a new-physics phase
$\theta_s$ is present in $\bs$-$\bsbar$ mixing, then it is not the
quantities ${\tilde Z}_R$ and ${\tilde Z}_I$ [Eqs.~(\ref{ZPR}) and
(\ref{ZPI})] which are measured, but rather $\tilde{Z}_R^{ex}$ and
$\tilde{Z}_I^{ex}$:
\bea
\tilde{Z}_R^{ex} &\equiv & Re\left( e^{-i\theta_s }A_s^{\phi*}
                                  \bar{A}^{\phi}_{s } \right) 
            =  \cos\theta_s \tilde{Z}_I - \sin\theta_s \tilde{Z}_R ~,
               \\
\tilde{Z}_I^{ex} &\equiv & Im\left( e^{-i\theta_s } A_s^{\phi*}
                                  \bar{A}^{\phi}_{s } \right) 
            =  \cos\theta_s \tilde{Z}_R + \sin\theta_s \tilde{Z}_I ~.
\eea
However, ${\tilde Z}_R$ and ${\tilde Z}_I$ can be obtained
straightforwardly:
\bea
\tilde{Z}_R &=&  \cos\theta_s \, \tilde{Z}_I^{ex} 
                + \sin\theta_s \, \tilde{Z}_R^{ex}
               \\
\tilde{Z}_I &=&  - \sin\theta_s \, \tilde{Z}_I^{ex} 
                 + \sin\theta_s \, \tilde{Z}_R^{ex}
\eea
Thus, assuming that $\theta_s$ is known independently (e.g.\ via any
of the methods we have described earlier), we can use these
expressions for ${\tilde Z}_R$ and ${\tilde Z}_I$ and simply apply the
above method. If $\theta_s$ is only known up to a discrete ambiguity,
then this simply increases the number of possible solutions for
$\beta$. However, in general we will still be able to determine that
$\beta \ne {\tilde\beta}$.

Finally, we note that even if there is no new physics (i.e.\ $\beta =
{\tilde\beta}$), this method still yields important information. If
one probes $\beta$ in the conventional way via CP violation in $\bd(t)
\to J/\Psi\ks$, one extracts the function $\sin 2\beta$. This gives
the angle $\beta$ up to a fourfold discrete ambiguity: if $\beta_0$ is
the true solution, $\beta_0 + \pi$, ${\pi\over 2} - \beta_0$ and
${3\pi\over 2} - \beta_0$ are also solutions for $\beta$. Our
technique can be used to eliminate two of these solutions. In
particular, due to the presence of the $\sin\beta$ and $\cos\beta$
factors in Eqs.~(\ref{XBSPK})-(\ref{ZBDKK}), ${\pi\over 2} - \beta_0$
and ${3\pi\over 2} - \beta_0$ will not in general be among the
solutions to these equations. However, $\beta_0 + \pi$ will still be
allowed if we simultaneously take $\Delta \to \Delta + \pi$ and
$\tilde\Delta \to \tilde\Delta + \pi$. Thus, in the absence of new
physics, the above method can be used to reduce the discrete ambiguity
in $\beta$ from a fourfold one to a twofold one.

\section{$\bs(t)\to J/\Psi \ks$ and $\bd(t)\to J/\Psi \pi^0$}

It is not difficult to think of variations on the above method. As a
second example, consider the decays $\bs\to J/\Psi \ks$ and $\bd\to
J/\Psi \pi^0$. Both of these decays get contributions from
colour-suppressed tree diagrams, Zweig-suppressed gluonic penguins and
electroweak penguins. If the penguins are not too small compared to
the tree diagram, it may be possible to extract $\beta$ and compare it
with ${\tilde\beta}$.

The amplitudes for these decays can be written as
\bea
A_s^{\psi} &\equiv &  A\left(B_s\rightarrow J/\psi K_s \right) =  
                    \tilde{C} + \tilde{P}_1 + \tilde{P}_{\sss EW} ~, \\
A_d^{\psi \pi } &\equiv &  A\left(B_d\rightarrow J/\psi {\pi }^0 \right) =  
                  C + P_1 + P_{\sss EW} ~,
\eea 
where $\tilde{C}$ and ${C}$ are the colour-suppressed tree amplitudes
in $\bs$ and $\bd$ decays, respectively. The combination of CKM matrix
elements involved in these amplitudes is $V_{cb}^* V_{cd}$, which is
real in the Wolfenstein parametrization.

As before, we use CKM unitarity to eliminate the $u$-quark piece of
the penguin contributions, allowing us to write
\bea
A_s^{\psi} &\equiv & \tilde{\cal C}e^{i {\tilde\delta}_C} +
             \tilde{\cal P}_{1} 
e^{i {\tilde\delta}_{{P}}} e^{-i\beta } ~, \\ 
A_d^{\psi \pi} &\equiv & {\cal C}e^{i \delta_C} +
             {\cal P}_{1} e^{i \delta_{{P}}} e^{-i\beta } ~,
\eea
where 
\bea
\tilde{\cal C} e^{i {\tilde\delta}_C} & = &
\tilde{C} + \tilde{P}_{1,c} - \tilde{P}_{1,u} ~, \nn\\
\tilde{\cal P}_1 e^{i {\tilde\delta}_P} e^{-i\beta } & = & 
\tilde{P}_{1,t} - \tilde{P}_{1,u} + \tilde{P}_{\sss EW} \nn\\
{\cal C} e^{i \delta_C} & = & C + P_{1,c} - P_{1,u} ~, \nn\\
{\cal P}_1 e^{i {\delta}_P} e^{-i\beta } & = & 
P_{1,t} - P_{1,u} + P_{\sss EW} ~.
\label{CPdefs}
\eea

In terms of these quantities, the time-dependent measurements yield
the following:
\bea
\tilde{X} &\equiv & \frac{1}{2}
            \left( 
                \left| A_s^{\psi}\right|^2 
              + \left| \bar{A}^{\psi}_s \right|^2 
            \right) = 
            \tilde{\cal C}^2 +\tilde{\cal P}_1^2 
      + 2 \tilde{\cal C} \tilde{\cal P}_1 \cos{\tilde\delta} \cos\beta ~,
\label{XBSJK} \\
\tilde{Y} &\equiv & \frac{1}{2}
            \left( 
                \left| A_s^{\psi}\right|^2 
              - \left| \bar{A}^{\psi}_s \right|^2 
            \right) =  - 2 \tilde{\cal C}
                      \tilde{\cal P}_1 \sin{\tilde\delta} 
                    \sin\beta ~, 
\label{YBSJK} \\
\tilde{Z}_R &\equiv & Re\left( A_s^{\psi*}
                                  \bar{A}^{\psi}_{s } \right) 
            =  - \tilde{\cal C}^2 - \tilde{\cal P}_1^2 \cos2\beta 
      - 2 \tilde{\cal C} \tilde{\cal P}_1 \cos{\tilde\delta} \cos\beta ~,
\label{ZRBSJK} \\
\tilde{Z}_I &\equiv & Im\left( A_s^{\psi*}
                                  \bar{A}^{\psi}_{s } \right) 
            =  - \tilde{\cal P}_1^2 \sin2\beta
      - 2 \tilde{\cal C} \tilde{\cal P}_1 \cos{\tilde\delta}
                   \sin\beta ~,
\label{ZIBSJK} \\
{X} &\equiv & \frac{1}{2}
            \left( 
                \left| A_d^{\psi \pi}\right|^2 
              + \left| \bar{A}^{\psi \pi }_d \right|^2 
            \right) = 
            {\cal C }^{2} +{\cal P}_1^{2} 
      + 2 {\cal C} {\cal P}_1 \cos\delta \cos\beta ~,
\label{XBDJP} \\
{Y} &\equiv & \frac{1}{2}
            \left( 
                \left| A_d^{\psi \pi}\right|^2 
              - \left| \bar{A}^{\psi \pi}_d \right|^2 
            \right) =  -  2 {\cal C}
                     {\cal P}_1 \sin\delta \sin\beta 
\label{YBDJP} \\
{Z} &\equiv & Im\left( e^{-2i \tilde{\beta }} A_d^{\psi \pi*}
                                  \bar{A}^{\psi \pi}_{d } \right) \nn \\
&~& \qquad\qquad  =  - {\cal C}^{2} \sin 2\tilde{\beta }  
            -  {\cal P}_1^{2} \sin(2 \tilde{\beta } - 2 \beta ) 
      - 2 {\cal C} {\cal P}_1 \cos\delta
\sin(2 \tilde{ \beta } - \beta ) ~,
\label{ZBDJP} 
\eea
where ${\tilde\delta} \equiv {\tilde\delta}_C - {\tilde\delta}_P$ and
$\delta \equiv \delta_C - \delta_P$.

Once again, this gives us six independent equations in seven unknowns.
However, we can reduce the number of parameters by assuming that $r =
\tilde{r}$, where $r \equiv {\cal P}_1 / {\cal C}$ and ${\tilde r}
\equiv \tilde{\cal P}_1 / \tilde{\cal C}$. Looking at
Eq.~(\ref{CPdefs}), it is clear that this assumption is well-justified
-- in fact, within the spectator model, the equality is exact. In this
case the observables become
\bea
\tilde{X} &=& \tilde{C}^2 
                [ 1 + 2 r \cos{\tilde\delta} \cos\beta + r^2 ]
\label{TXBSJK} \\
\tilde{Y} &=& \tilde{\cal C}^2 
                [ - 2 r \sin{\tilde\delta} \sin\beta ]
\label{TYBSJK} \\
\tilde{Z}_R &=& \tilde{\cal C}^2 
                [ - 1 - 2r \cos{\tilde\delta}\cos\beta - r^2 \cos2\beta ]
\label{TZRBSJK} \\
\tilde{Z}_I &=& \tilde{\cal C}^2 
                [ - 2r \cos{\tilde\delta}\sin\beta - r^2 \sin2\beta ]
\label{TZIBSJK} \\
X &=& {\cal C}^2 [ 1 + 2 r \cos{\delta} \cos\beta + r^2 ]
\label{XDJP} \\
Y &=& {\cal C}^2 [ -  2 r \sin{\delta} \sin\beta ]
\label{YDJP} \\
Z &=& {\cal C}^2 [ - \sin2\tilde{\beta } - 
                    2 r \cos\delta \sin( 2\tilde{\beta} - \beta )
                  - r^2 \sin( 2\tilde{\beta} - 2\beta ) ] 
\label{ZDJP}
\eea
The form of these equations is similar to that found for
$\bs\to\phi\ks$ and $\bd \to K^0 \kbar$ decays
[Eqs.~(\ref{XBSPK})-(\ref{ZBDKK})]. As in that case, assuming that
${\tilde\beta}$ is measured in $\bd(t) \to J/\psi \ks$, we now have
six independent equations in six unknowns. And as before, it is
possible to solve these equations for the six parameters.  We can thus
obtain $\beta$, and test whether $\beta = {\tilde\beta}$ or not.

It must be admitted, however, that from a theoretical point of view
this method is much less compelling than the one which uses the decays
$\bs\to\phi\ks$ and $\bd \to K^0 \kbar$. We have noted that the
assumption that $r = \tilde{r}$ is well justified since the only
difference between the decays $\bs\to J/\Psi \ks$ and $\bd\to J/\Psi
\pi^0$ is the flavour of the spectator quark. However, this is also
problematic: if the flavour of the spectator quark is completely
irrelevant, then we also have ${\cal C} = \tilde{\cal C}$ and $\delta
= \tilde\delta$. But in this case the $\bd\to J/\Psi \pi^0$ decay
gives us no extra information: we have $X = \tilde{X}$, $Y =
\tilde{Y}$, and $Z = \tilde{Z}_R \cos 2{\tilde\beta} - \tilde{Z}_I
\sin 2{\tilde\beta}$.  So we are back to the situation of having three
equations in four unknowns, which obviously cannot be solved.

Thus, for this method to work, not only is it necessary for the
penguin contributions to $\bs\to J/\Psi \ks$ and $\bd\to J/\Psi \pi^0$
to be sizeable, there must also be significant differences in the
sizes of the contributing amplitudes due to the flavour of the
spectator quark. While it is possible that these conditions are
fulfilled, it is theoretically disfavoured. Thus, the method involving
the decays $\bs\to\phi\ks$ and $\bd \to K^0 \kbar$ is probably more
promising.

\section{Conclusions}

CP-violating asymmetries in $B$ decays will be measured in the near
future. If Nature is kind, we will find evidence of physics beyond the
SM. The most obvious signal of new physics will be if the unitarity
triangle constructed from measurements of the CP angles disagrees with
that constructed from independent measurements of the sides. The
problem here is that there are significant theoretical uncertainties
in the measurements of the sides. Thus, even if there is a
discrepancy, it may not provide compelling evidence for new physics --
it may simply be that the errors on the theoretical parameters have
been underestimated. For this reason, it is important to find ways of
directly probing for new physics in the $B$ system.

Although there are several methods which will allow us to directly
test for the presence of new physics in the $b \to s$ FCNC (either in
$\bs$-$\bsbar$ mixing or in $b \to s$ penguins), finding new physics
in the $b\to d$ FCNC ($\bd$-$\bdbar$ mixing or $b\to d$ penguins) is
considerably more difficult. To date, no methods have been suggested
which directly probe new physics in the $b\to d$ FCNC.

In this paper we have discussed a method to search for new physics in
the $b\to d$ FCNC. By making time-dependent measurements of the decays
$\bs(t)\to\phi\ks$ and $\bd(t) \to K^0 \kbar$ it is possible to
compare the weak phase in $\bd$-$\bdbar$ mixing with that of the
$t$-quark contribution to the $b\to d$ penguin. Since these phases are
equal in the SM, any discrepancy would be a clear signal of new
physics. The method is not entirely free of hadronic uncertainties: it
does require some theoretical input. Still, the necessary assumption
--- that a ratio of amplitudes in the two decays is equal --- is
reasonably well-justified theoretically. We estimate the error in this
assumption to be $\lsim 20\%$.

This method can be applied to other decay modes. For example, we have
also examined the decays $\bs(t)\to J/\Psi \ks$ and $\bd(t)\to J/\Psi
\pi^0$. Although in principle new physics in the $b\to d$ FCNC can be
found using this set of decays, the necessary conditions are
theoretically disfavoured. Thus the decays $\bs(t)\to\phi\ks$ and
$\bd(t) \to K^0 \kbar$ are more promising.

\section*{\bf Acknowledgments}

C.S.K. wishes to acknowledge the financial support of 1997-sughak program
of Korean Research Foundation. The work of D.L.
was financially supported by NSERC of Canada and FCAR du Qu\'ebec. The
work of T.Y. was supported in part by Grant-in-Aid for Scientific
Research from the Ministry of Education, Science and Culture of Japan
and in part by JSPS Research Fellowships for Young Scientists.

\end{document}